\documentclass{PoS}

\usepackage[utf8]{inputenc}
\usepackage{graphicx}
\usepackage{dcolumn}
\usepackage{bm}
\usepackage{epstopdf}
\usepackage{mathrsfs}
\usepackage{amssymb,amsmath,amsfonts,latexsym}
\usepackage{nicefrac}
\usepackage{tikz}

\usepackage{lipsum}
\usepackage{nicefrac}
\usepackage[normalem]{ulem}

\def\0{{\boldsymbol 0}}

\def\Q{{Q}}
\def\C{{C}}
\def\P{{P}}

\def\med{\text{med}}

\def\jet{\text{jet}}

\def\tform{{t_\text{f}}}
\def\tdecoh{t_\text{d}}

\def\pT{p_{\sst T}}

\def\pT{p_{_T}}

\newcommand{\beq}{\begin{eqnarray}}
\newcommand{\eeq}{\end{eqnarray}}
\newcommand{\be}{\begin{eqnarray*}}
\newcommand{\ee}{\end{eqnarray*}}
\newcommand{\bal}{\begin{align}}
\newcommand{\eal}{\end{align}}
\newcommand{\rmd}{{\rm d}}
\newcommand{\dd}{{\rm d}}
\newcommand{\rme}{{\rm e}}

\def\abar{{\rm \bar\alpha}}
\newcommand{\eqn}[1]{Eq.~\eqref{#1}}
\newcommand{\nn}{\nonumber\\ }

\title{Higher-order corrections to jet quenching}

\ShortTitle{Higher-order corrections to jet quenching}

\author{\speaker{Yacine Mehtar-Tani}\\
        Brookhaven National Laboratory, Physics Department, Upton, NY 11973, United States\\
        E-mail: \email{mehtartani@bnl.gov}}
\author{Konrad Tywoniuk\\
        University of Bergen, Postboks 7803, 5020 Bergen, Norway\\
      E-mail: \email{konrad.tywoniuk@uib.no}}

\abstract{ We compute the inclusive jet spectrum in the presence of a dense QCD medium by going beyond the single parton energy loss approximation. We show that higher-order corrections are important yielding large logarithmic contributions that must be resummed to all orders. This reflects the fact that jet quenching is sensitive to fluctuations of the jet substructure. }

\FullConference{International Conference on Hard and Electromagnetic Probes of High-Energy Nuclear Collisions\\
		30 September - 5 October 2018\\
		Aix-Les-Bains, Savoie, France}

\begin{document}

\section{Introduction}
The phenomenon of jet quenching is regarded as one of the strongest evidence for the formation of a hot and dense de-confined matter in ultra-relativistic heavy ion collisions, the so-called quark-gluon plasma (QGP). Final state interactions cause high-$p_T$ jets to lose a substantial amount of their energy to the QGP via elastic and inelastic processes, and to migrate to lower $p_T$ bins resulting in an overall suppression of the inclusive jet spectrum. Remarkably, this strong suppression persists over a large range of transverse jet momentum up to the TeV scale \cite{Aaboud:2018twu}. 
The standard analytic approaches to jet quenching are based on single parton radiative energy loss in which hard color charges lose energy by radiating soft gluons at large angles w.r.t. the jet axis (see recent reviews \cite{Mehtar-Tani:2013pia,Blaizot:2015lma} and references therein). 
However, owing to the QCD mass singularity, the jet-initiating parton tends to branch rapidly. As a result, the jet fluctuates into many color charges very early on inside the medium, and energy loss is expected to scale with the number of jet constituents. To gauge the importance of such early splittings we estimate the phase-space for a highly virtual quark to branch inside the medium. Requiring that the quantum mechanical formation time $t_f = \omega/k_\perp^2$ (where $\omega$ and $k_\perp$ are the energy and  transverse momentum of the radiated gluon during the splitting)  be smaller than the length of the medium $L$ one finds 
\beq \label{eq:PS-naive}
\text{PS}\equiv  \abar \int_0^{p_T} \frac{\rmd \omega }{\omega }\int_0^R \frac{\rmd \theta}{\theta}\, \Theta(t_f  < L)=\frac{\abar}{2}\, \ln^2 p_T R^2 L,
\eeq
where $p_T$, $R$ are the jet opening angle. 
For $L=5$ fm, $R=0.3$ rad, $p_T=500$ GeV and $\abar \equiv \alpha_s C_F/\pi= 0.2$, we find $\text{PS} \simeq 2.5 \gtrsim 1$. It follows that the probability of branching inside the medium is of order one. 

In vacuum, the jet $p_T$ is not sensitive to substructure fluctuations. This is reflected by strong cancellations between real and virtual contributions that is a consequence of unitarity.  However, in the presence of a medium this is no longer true as a real emission is associated with the process $n\to n+1$ partons whose energy loss is larger than the corresponding virtual fluctuation that does not affect the number of charges, i.e., $n \to n$. The mismatch arises because energy loss increases with the number of color charges. It has been pointed out recently in Monte Carlo studies that fluctuations related to the jet substructure are  important for understanding experimental data, see e.g. \cite{Milhano:2015mng,Casalderrey-Solana:2016jvj}. However, we expect interference effects, that are neglected in most MC implementations, to play a role in jet quenching  \cite{Mehtar-Tani:2017ypq,MehtarTani:2010ma,MehtarTani:2011tz,CasalderreySolana:2011rz}.  

\section{Phase-space of in-medium vacuum-like cascade}

\begin{figure}
\includegraphics[width=0.4\textwidth]{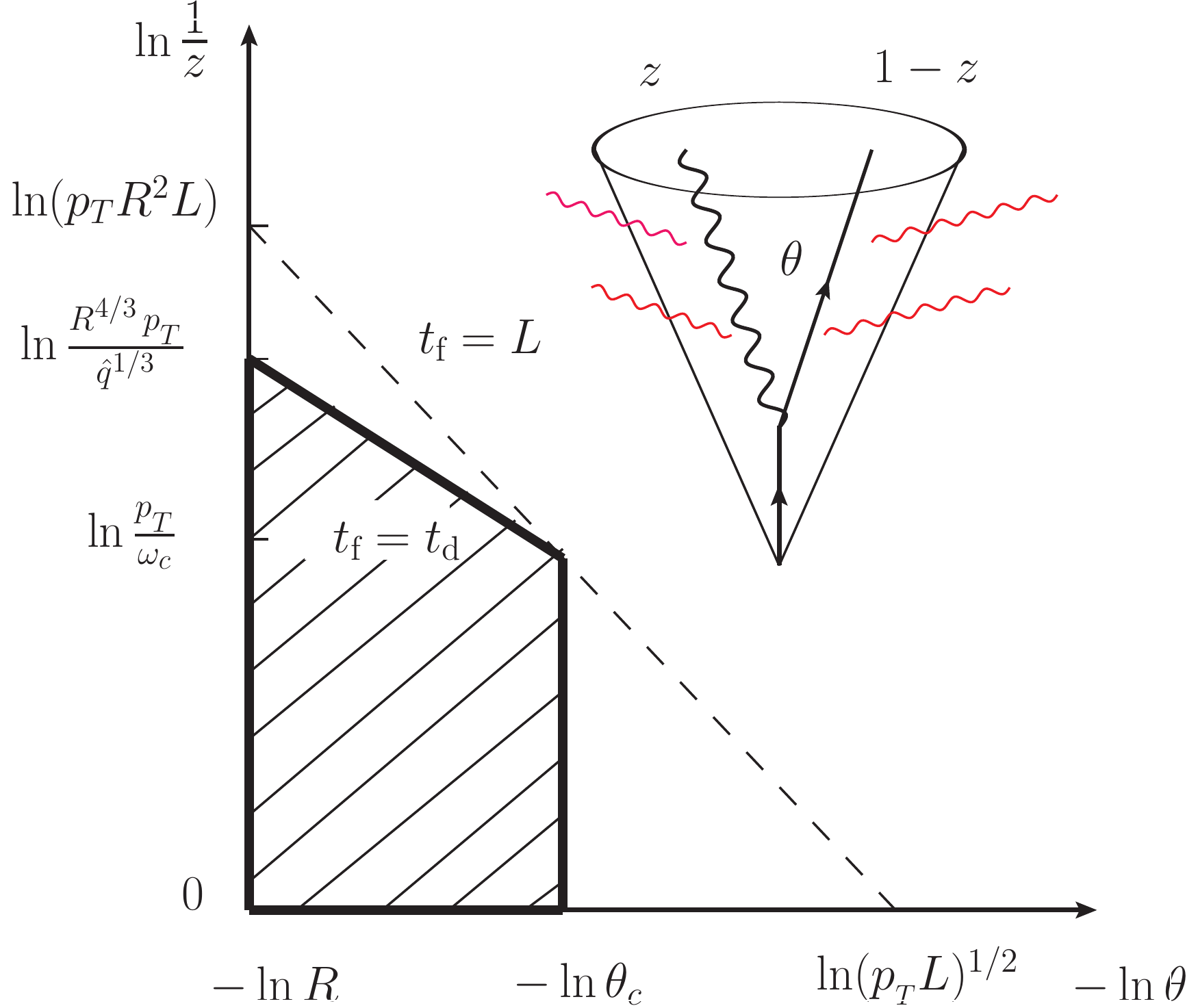}\qquad\qquad\includegraphics[width=0.45\textwidth]{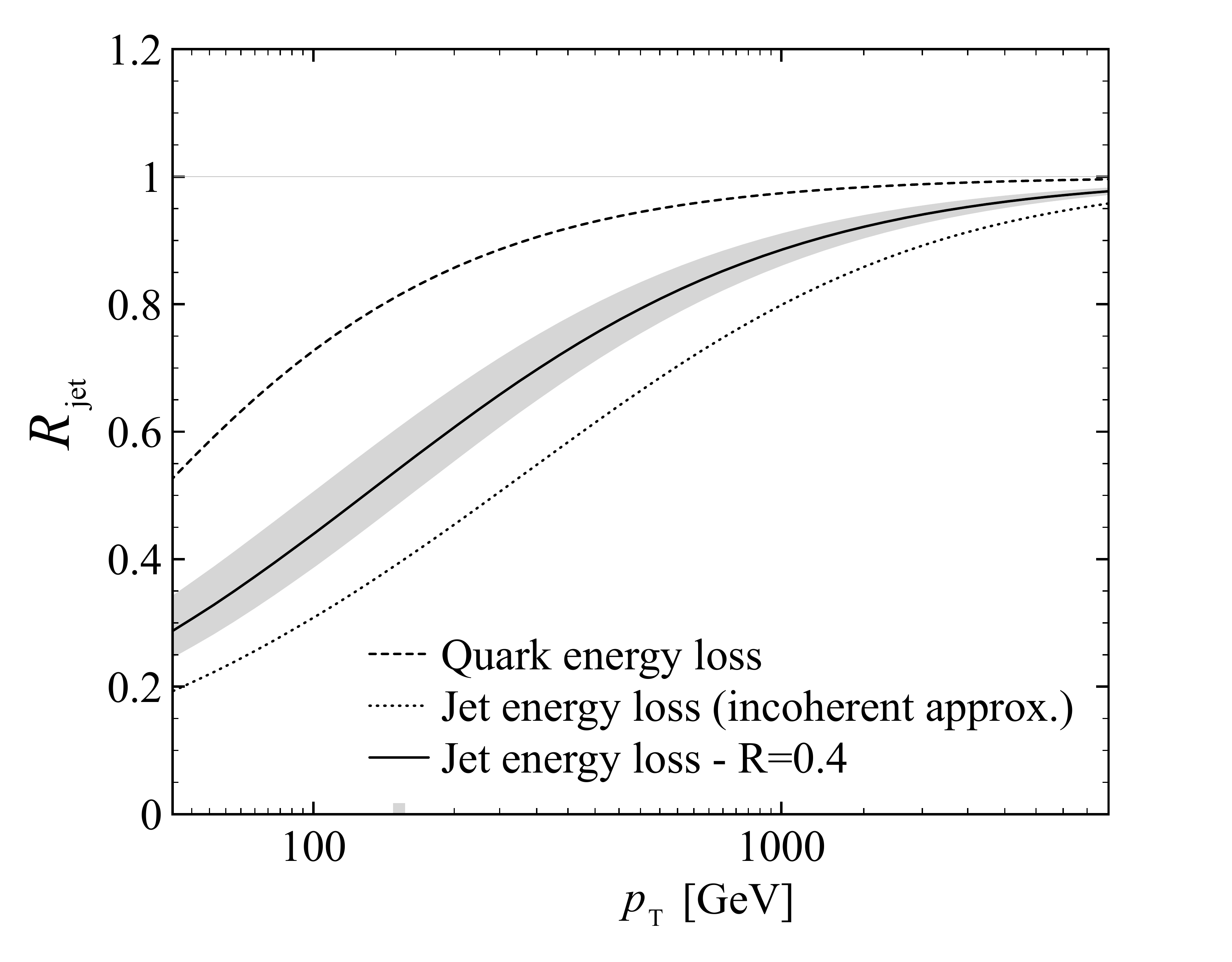}
\caption{Left: The phase-space in the $(\ln(1/z)-\ln(1/\theta))$ plane, where $z=\omega/p_T$, of higher-order corrections to the jet inclusive spectrum in the presence of radiative energy loss depicted by the shaded region. We have included lines denoting $\tform = L$, $\tform = \tdecoh$ and $\tdecoh = L$ (implying $\theta = \theta_c$). Right: Quark jet suppression factor calculated using single parton quenching weight alone (dashed line) and additionally supplemented by the Sudakov suppression with (solid) and without (dotted line) coherence effects.}
\label{fig:nloeloss}
\end{figure}

The mismatch between real and virtual contributions takes place whenever the medium resolves the individual color charges produced in hard splittings. This occurs at the decoherence time $t = \tdecoh \sim (\hat q \theta^2)^{-1/3}$, when the transverse distance between the two daughters, $r_\perp\sim \theta  t$, where $\theta$ (in rad) is the angle formed by them, becomes of order the medium correlation length $(\hat q t)^{-1/2}\sim k_\perp^{-1}$.  Here, $\hat q \equiv \rmd \langle k^2_\perp\rangle /\rmd t$, the so-called quenching parameter, is the typical transverse momentum squared acquired by a parton in the medium per units time $t$. 
This corresponds to a minimum characteristic angle $\theta_c \sim (\hat q L^3)^{-1/2}$ (when $\tdecoh = L$), below which the jet is on average unresolved by the medium, i.e., when $R < \theta_c$. In this case, the energy loss probability distribution is that of the total charge, namely the parent parton, and hence, is not sensitive to the fluctuations of the jet substructure.
Accounting for this constraint, the phase-space for higher order corrections is depicted in Fig.~\ref{fig:nloeloss} (left) where the shaded area represents the region in $(\omega,\theta)$ plane corresponding to vacuum-like splittings that take place inside the medium and that are resolved by it, in terms of constraint on the formation time we have $t_f \ll t_d \ll L$ \cite{Mehtar-Tani:2017web,Caucal:2018dla}. 

\beq
\label{eq:PhaseSpace}
\text{PS}\equiv 
2\abar \int_0^R  \frac{\rmd \theta}{\theta} \int_0^{p_T} \frac{\rmd \omega}{\omega} \,\Theta(\tform < \tdecoh<L)= 2\abar \ln \frac{R}{\theta_c } \left (\ln \frac{\pT}{\omega_c}  + \frac{2}{3} \ln \frac{R}{\theta_c } \right) \,,
\eeq
where $\omega_c=\hat q L^2$ is the maximum medium-induced frequency. Note that due to color coherence the relevant phase-space for real emissions that are resolved by the medium is smaller than the naive expectation Eq.~(\ref{eq:PS-naive}) that corresponds the triangle formed by the dashed line and the plot axes. 

\section{Jet nuclear modification factor }

The amount of energy, $\epsilon$, radiated out of the jet cone is generically described by the probability distribution $\P(\epsilon)$, called the quenching weight \cite{Baier:2001yt,Salgado:2003gb}, 
\beq
\label{eq:spect-general}
\frac{\rmd \sigma_\med }{\rmd \pT^2\rmd y} = \int^\infty_0 \rmd \epsilon\, \P(\epsilon) \, \frac{\rmd \sigma_\text{vac} (\pT+\epsilon)}{\rmd\pT^2 \rmd y} \,,
\eeq
where $\rmd \sigma_\text{vac} $ is the jet spectrum in vacuum, that is, in the absence of final state interactions . 
The quenching probability distribution depends implicitly on the medium properties, such as the jet quenching parameter $\hat q$, which corresponds to the in-medium diffusion coefficient in transverse momentum space,  and the medium length $L$, as well as  jet $\pT$ and cone size $R$.

In what follow, we shall focus on the nuclear modification factor defined as follows:
\beq
R_\jet \equiv Q(\pT)=\frac{\rmd  \sigma_\med (\pT)}{\rmd\pT^2 \rmd y}\Big/ \frac{\rmd \sigma_\text{vac} (\pT)}{\rmd\pT^2 \rmd y}. 
\eeq
One expects different jet configurations to lose a different amount of energy and hence, $\P(\epsilon)$ and in turn $Q(\pT)$ should be sensitive to the fluctuations of the jet substructure. If one assumes a steeply falling spectrum with a constant power index $n$, one can make the following useful approximation \cite{Baier:2001yt}
$\rmd \sigma_\text{vac} (\pT+\epsilon)/\rmd \pT^2 \rmd y  \approx \rmd \sigma_\text{vac}/ \rmd \pT^2 \rmd y \,  \exp \left(-n \epsilon/p_{_T}\right)$,
which holds for small $\epsilon/\pT$ and large $n$. Using this form, one observes that the quenching factor is related to the Laplace transform of the quenching weight, $\tilde \P (\nu) = \int_0^\infty\dd \epsilon \,\P(\epsilon) \,\rme^{-\nu \epsilon}$, as follows $R_\text{jet} =\Q(\pT)$ where $\Q(\pT) \equiv \tilde \P\left(n\big/\pT\right)$.  The idea behind this approximation is to turn the convolution of multiple energy loss probabilities, when multiple partons are involved at higher orders, into a direct product of single parton quenching factors. The  quenching factor can be expanded in powers of the strong coupling constant $\Q(\pT) = \Q^{(0)}(\pT) + \Q^{(1)}(\pT) +O(\alpha_s^2)$, where the leading order corresponds to the energy loss of the parent parton: 
$\Q_q(\pT) = \exp\left[\int_0^\infty \rmd \omega \, \frac{\rmd I}{\rmd \omega}\,(\rme^{- n \omega / \pT}-1)\right] \approx \exp[- 2 \abar L \sqrt{\pi n \hat q/\pT}] $ where the medium-induced radiation spectrum is given by $\omega \rmd I/\rmd \omega \approx \abar \sqrt{\omega_c /\omega} $ \cite{Baier:1996sk,Zakharov:1997uu} in the multiple-soft scattering approximation. Limiting our discussing the dominant logarithmic contribution in the large $N_c$ limit,  the first corrections reads
\beq
\!\Q^{(1)}( \pT) = \frac{\alpha_s}{\pi}\!\!\int_0^1 \!\rmd z \, P_{gq}(z) \!\int_0^R \frac{\rmd \theta}{\theta}\,  \,\Theta(t_f<t_\text{d}<L)\,  \Big( \Q_q^2(\pT) -1\Big)\Q_R( \pT) \,
\label{eq:nlo-quenching}
\eeq
where the integrations over $z=\omega/\pT$ and $\theta$ are confined to the phase-space region \eqn{eq:PhaseSpace}.
the first and second terms correspond to  the real and virtual  contributions respectively. Note that, in addition to the total charge quenching factor, $\Q_R(\pT)$ where $R=q,g$ is the representation of the parent parton, the real contribution received an additional suppression factor that corresponds to the quenching of the quark dipole that is associated with the radiation of a gluon in the large-$N_c$ limit.  Most importantly it reflects the fact that the presence of additional color charges yield more suppression. 
Iterating the above formula to all order using the coherent branching algorithm \cite{Dokshitzer:1991wu}, that is, assuming strict angular ordering along the shower we can write an non-linear evolution equation for the following quantity 
\beq
C(\pT) \equiv \frac{\Q(\pT)}{\Q_R(\pT)},
\eeq
which is a measure of jet-quenching effect of jet substructure fluctuations \cite{Mehtar-Tani:2017web}:

\begin{align}
\label{eq:collimator-evol-2}
\C_i (1,\pT,R)  &= 1 + \int_0^1 \rmd z  \int_{\theta_c}^R \frac{\rmd \theta}{\theta} \,\frac{\alpha_s(k_{\perp}) }{\pi}P_{gi}(z)\Theta(\tdecoh-\tform )
 \nn
&\times \Big[ \C_g (z,\pT,\theta) \C_i \big((1-z),\pT,\theta\big) \Q_q^2(\pT) - \C_i(1,\pT,\theta) \Big] \,,
\end{align}
where $i=q,g$ and we have restored the full Altarelli-Parisi splitting functions (cf. $P_{gq}(z) = C_F(1+(1-z)^2)^2/[z(1-z)]$). 
$C(\pT)=1 $ implies that the nuclear modification factor is given solely by the quenching factor of the total charge $\Q_R(\pT)$.  
It is instructive to analyze the limit of strong quenching, i.e., $\Q_q(\pT) \ll 1$, for $\pT \ll n \abar^2 \hat q L^2 $. In this case, one can neglect the non-linear term.  As a result one obtains the exponentiation of the phase-space \eqn{eq:PhaseSpace} into the Sudakov form factor \cite{Mehtar-Tani:2017web},
\beq
\label{eq:collimator-evol-final1}
\C_q(\pT,R)\simeq \exp\left[-2 \abar  \ln \frac{R}{\theta_c } \left (\ln \frac{p_{_T}}{\omega_c}  + \frac{2}{3} \ln \frac{R}{\theta_c } \right)\right],
\eeq 
when $\pT>\omega_c$, where we have only taken into account the soft limit of the splitting function. 
The numerical solution of \eqn{eq:collimator-evol-2} is shown in Fig.~\ref{fig:nloeloss} (right) for $\hat q=1$ GeV$^2$/fm and $L=3$ fm. The dashed and dotted curves correspond to the cases where the jet is approximated by a single parton and where color coherence is neglected, respectively. This demonstrates the sensitivity of the nuclear modification factor to interference effects in energy loss.
\section{Discussion}
We have shown that the inclusive jet spectrum is sensitive to fluctuations of the jet substructure and in particular to color coherence that reduces the phase space for resolved jet color charges by the medium. For more quantitative estimates we plan to include in a future work in-cone medium radiation that we have neglected in the present study, as well as momentum broadening and total quenching of the jet constituents. 
\section*{Acknowledgements} 
YMT is supported by the U.S. Department of Energy, Office of Science, Office of Nuclear Physics, under contract No. DE- SC0012704, and in part by Laboratory Directed Research and Development (LDRD) funds from Brookhaven Science Associates. KT is supported by a Bergen Research Foundation (BFS) Starting Grant “Thermalising jets: novel aspects of non-equilibrium processes at colliders”.

\end{document}